\documentclass[aps,prb,twocolumn,superscriptaddress,showpacs,floatfix]{revtex4}
\usepackage{graphicx}
\usepackage{latexsym}
\usepackage{stmaryrd}
\usepackage{amsmath}
\usepackage{amssymb}
\usepackage{epstopdf}
\usepackage[colorlinks,linkcolor=red,anchorcolor=red,citecolor=blue]{hyperref}

\begin{document}
\makeatletter
\newcommand{\rmnum}[1]{\romannumeral #1}
\newcommand{\Rmnum}[1]{\expandafter\@slowromancap\romannumeral #1@}
\makeatother

\title{The Global Phase Diagram of disordered Higher-order Weyl Semimetals}

\author{Zhi-Qiang Zhang }
\affiliation{School of Physical Science and Technology, Soochow University, Suzhou, 215006, China}
\author{Bing-Lan Wu }
\affiliation{School of Physical Science and Technology, Soochow University, Suzhou, 215006, China}
\author{Chui-Zhen Chen }
\affiliation{Institute for Advanced Study, Soochow University, Suzhou 215006, China}
\affiliation{School of Physical Science and Technology, Soochow University, Suzhou, 215006, China}
\author{Hua Jiang}\email{jianghuaphy@suda.edu.cn}
\affiliation{School of Physical Science and Technology, Soochow University, Suzhou, 215006, China}
\affiliation{Institute for Advanced Study, Soochow University, Suzhou 215006, China}
\date{\today}

\begin{abstract}
We study the disorder-induced phase transition of higher-order Weyl semimetals (HOWSMs) and the fate of the topological features of disordered HOWSMs. We obtain a global phase diagram of HOWSMs according to the scaling theory of Anderson localization.
 Specifically, a phase transition from the Weyl semimetal (WSM) to the HOWSM is uncovered, distinguishing the disordered HOWSMs from the traditional WSMs. Further, we confirm the robustness of Weyl-nodes for HOWSMs.
 Interestingly, the unique topological properties of HOWSMs show different behaviors: (i) the quantized quadrupole moment and the corresponding quantized charge of hinge states are fragile to weak disorder; (ii) the hinge states show moderate stability which enables the feasibility in experimental observation. Our study deepens the understanding of the topological nature of HOWSMs and paves a possible way to the characterization of such a phase in experiments.
\end{abstract}

\maketitle

\section{Introduction}

 Over the past decades, topological states have become an important and blooming research area in condensed matter physics.  A large number of topological phases are theoretically predicted and experimentally observed \cite{TI1,TI2,WSM1,WSM2,HOTI1,HOTI2,HOTI3,HOTI4,HOTI5}, among which the higher-order topological states attract great interests in recent years\cite{HOTI1,HOTI2,HOTI3,HOTI4,HOTI5,HOTSC1,HOTSC2,HOTSC3,HOWSM1,HOWSM2,HOWSM3,HOWSM4}.
 A $n$-dimensional higher-order topological phase shows its unique topological features in $(n-d)$ dimension with $d\ge2$.
  For example, the two (three) dimensional higher-order topological insulators have corner (hinge) states, which are the most focused nowadays.
 Furthermore, a new kind of topological semimetal with Weyl-nodes, which is also named higher-order Weyl semimetal (HOWSM)\cite{HOWSM1,HOWSM2,HOWSM3,HOWSM4}, gains special attention very recently. The HOWSMs are predicted to possess both the Fermi-arc surface states and hinge states\cite{HOWSM1,HOWSM2,HOWSM3}. In particular, the hinge states hold the quantized charge as well as the quantized electric quadrupole moment. These features make HOWSMs different from the traditional Weyl semimetals (WSMs)\cite{WSM1,WSM2}. Such topological states are also reported to have been realized in classical-wave systems experimentally\cite{HOWSM4,HOWSM5}.

In order to comprehensively understand the topological properties of higher-order topological phases, many studies are concentrating on the disorder effect of these systems \cite{HOD1,HOD2,HOD3,HOD4,HOD5,HOD6,HOD7,HOD8}. In 2019,
  Araki {\it et al.}\cite{HOD1} used the machine learning method to study the phase transition of two-dimensional higher-order topological insulators. Kang {\it et al.}\cite{HOD2} and Wheeler {\it et al.}\cite{HOD3} compared the formula of the electric dipole moment and the electric quadrupole moment $Q_{xy}$ for higher-order topological insulators, and found that $Q_{xy}$ can be calculated in real space. However, since the system's symmetry may not preserve when disorder appears, the quantized $Q_{xy}$ is not always robust against disorder. In 2020, Li {\it et al.}\cite{HOD4} realized that the quantized quadrupole moment can still hold for systems with chiral symmetry or particle-hole symmetry. They consider a specific disorder scheme in the two-dimensional Su-Schrieffer-Heeger model, which ensures the chiral symmetry, and the higher-order topological Anderson insulator is obtained. Similar results are also achieved independently by Yang {\it et al.}\cite{HOD5}. Almost at the same time, the realization of the higher-order topological Anderson insulator in the electric circuit is also reported\cite{HOD6}.

Although there have been several works on studying two-dimensional systems, the investigations on the disorder effect of higher-order topological states in three-dimension are still lacking\cite{3DHOD1,3DHOD2,dirtyHOSM}.
Especially when the Weyl-nodes replace the insulator gap, the influence of disorder to the corresponding higher-order topological states is also unclear.
Thus, it is in urgent need to study the combination of disorder effect and HOWSMs,
which is essential to uncover the topological properties and guide the experimental observations of such phases.

\begin{figure*}[t]
   \centering
    \includegraphics[width=0.95\textwidth]{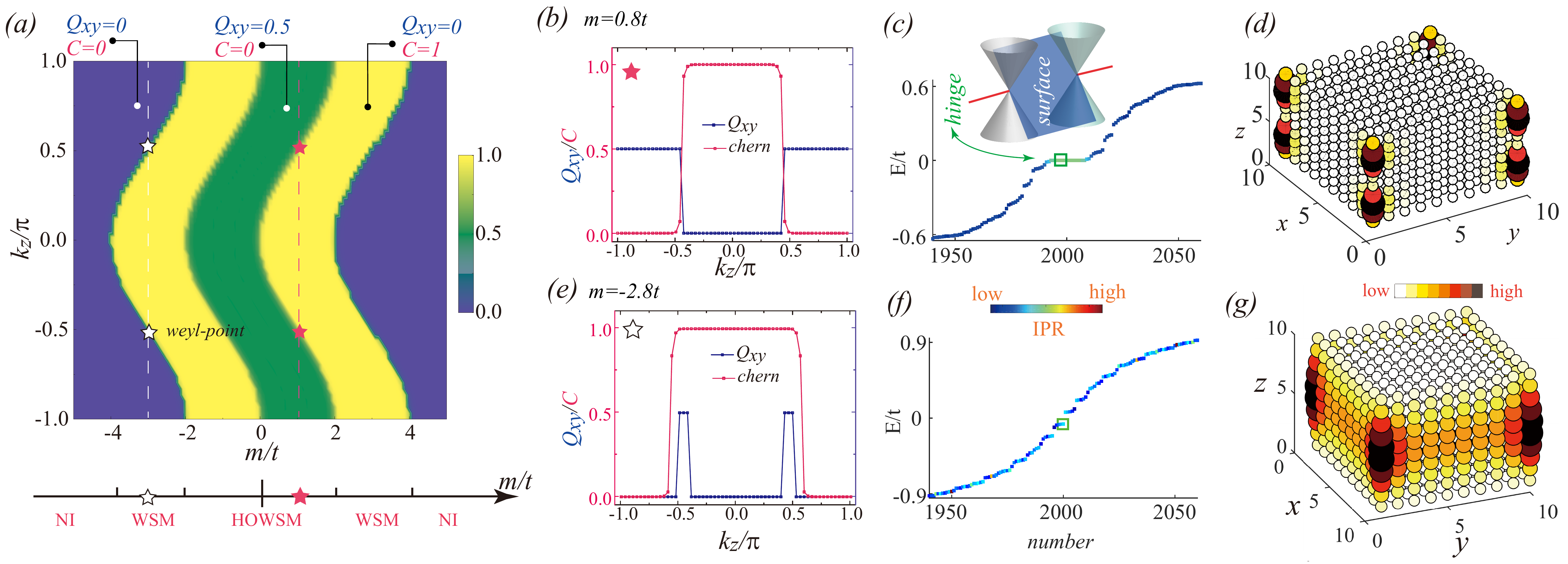}
    \caption{(Color online).
    (a) $\mathcal{C}+Q_{xy}\theta(1-\mathcal{C})$ versus momentum $k_z$ and topological mass $m$. $\mathcal{C}$ and $Q_{xy}$ are the Chern number and quadrupole moment, respectively. The green, yellow and blue regions correspond to $(\mathcal{C}=0, Q_{xy}=0.5)$, $(\mathcal{C}=1, Q_{xy}=0)$, and $(\mathcal{C}=0, Q_{xy}=0)$, respectively.
    There are three phases by increasing $m$, which are normal insulator (NI), WSM, and HOWSM.
     (b)-(d) belong to the HOWSM marked by the red star in (a).
     (b) $\mathcal{C}$ and $Q_{xy}$ versus $k_z$.
     (c) Counting the eigenvalue of a cubic sample with size $N_x\times N_y\times N_z=10\times 10\times10$. The color represents the IPR of eigenstates. The inset is a schematic diagram of the hinge states (solid red line) and the surface states (surface in blue) of HOWSMs. The hinge states correspond to the eigenstates with $E\approx0$.
     (d) Typical eigenstate distribution of the hinge states.
     (e)-(g) is the same with (b)-(d), except that they belong to the WSM marked by the white star in (a).   }
   \label{f1}
\end{figure*}

In this paper, we study the disorder-induced phase transition in HOWSMs. By investigating a three dimensional model with only one pair of Weyl-nodes, we pay attention to the stability of the HOWSMs in two aspects: (1) the stability of the Weyl-nodes of HOWSMs, (2) the fate of the unique topological features of HOWSMs [i.e., (i) the quantized quadrupole moment, (ii) the quantized charge of each hinge, and (iii) the hinge states].
  According to the finite-size scaling analysis, the Weyl-nodes of HOWSMs are robust against disorder. We also find disorder-induced phase transitions between HOWSMs and WSMs.
  However, the unique topological features of HOWSMs are not as robust as the Weyl-nodes.
   Specifically, the quantized quadrupole moment and the corresponding quantized electron charge of HOWSMs will be destroyed even under extremely small disorder strength. The hinge states are more robust and may be detectable in condensed matter experiments. However, its stability is still weaker than the Weyl nodes.

The rest of this paper is organized as follows: In Sec.~\ref{section2}, we present the model and the numerical methods. In Sec.~\ref{section3}, details of the model and the topological features of clean HOWSMs are shown. In Sec.~\ref{section4}, we study the disorder effect of HOWSMs. A global phase diagram and the fate of the unique topological features of disordered HOWSMs are obtained. Finally, a brief discussion and summary
is presented in Sec.~\ref{section5}.

\section{model and methods}\label{section2}

The minimum tight-binding Hamiltonian for the higher-order Weyl semimetal in cubic lattice reads\cite{HOWSM1}:
\begin{align}
\begin{split}
\mathcal {H}&= [m+t_x\cos k_x+t_y\cos k_y+t_z \cos k_z]\tau_z\sigma_0\\
&+B\tau_0\sigma_z+\lambda \sin k_x\tau_x\sigma_x+\lambda \sin k_y\tau_x\sigma_y\\
&+\Delta (\cos k_x-\cos k_y)\tau_y\sigma_0.
\label{EQ1}
\end{split}
\end{align}
$\tau_i$ and $\sigma_i$ are the pauli pauli matrices
 for orbit and spin basis with $i\in x,y,z$. $\sigma_0/\tau_0$ is the identity matrix. $\Delta$ is similar to the order parameter of $d$-wave superconductors, which has been widely used in higher-order topological superconductors\cite{HOTSC1}.
The second term describes the Zeeman-effect with splitting strength $B$, which is necessary to realize Weyl semimetals with only one pair of Weyl nodes\cite{WSMD1,WSMD2}.
For simplicity, we set parameters $t_x=t_y=t_z=t$, $\lambda=t$, $B=t$ and $\Delta=t$ throughout this paper. The disorder effect is introduced by the on-site potential $\epsilon_n $ with $n$ the lattice site. Here, $\epsilon_n\in[-W/2,W/2]$ corresponds to the Anderson disorder, and $W$ is the disorder strength\cite{WSMD1,WSMD2,WSMD3,WSMD4,WSMD5,WSMD6}.

Previous studies have shown that the Chern number $\mathcal{C}$ and quadrupole moment $Q_{xy}$ are important for HOWSMs\cite{HOWSM2,HOWSM3}. In our numerical calculation, $\mathcal{C}$ and $Q_{xy}$ defined in the real space are adopted.
The Chern number is calculated based on the non-commutative geometry method\cite{HOD8,chern1,chern2} with
\begin{equation}
\mathcal{C}=-\frac{2\pi i}{N_xN_y}\sum_{n,\alpha}\langle n,\alpha|P_r[-i[\widehat{x},P_r],-i[\widehat{y},P_r]] |n,\alpha\rangle.
\label{EQ2}
\end{equation}
$|n,\alpha\rangle$ represents the state sitting at position $n$ with orbit $\alpha$. $P_r=UU^\dagger$ is the projection operator of the occupied states, and $U$ is a matrix constructed by the occupied eigenvalues. $N_x$ ($\widehat{x}$) and $N_y$ ($\widehat{y}$) are the sample size (the coordinate operator) along $x$ and $y$ directions, respectively. Take $\widehat{x}$ as an example, the commutation relations can be obtained as follows:
\begin{equation}
[\widehat{x},P_r]=i\sum_{l=1}^{Q}c_l(e^{-il\widehat{x}\delta_{x}} P_re^{il\widehat{x}\delta_{x}}- e^{il\widehat{x}\delta_{x}} P_re^{-il\widehat{x}\delta_{x}}),
\end{equation}
with $\delta_x=2\pi/N_x$. Here, $Q$ is an integer, which determines the accuracy of the final results. For larger $Q$, the corresponding Chern number will be more accurate. $c_l$ is obtained by solving
\begin{equation}
 x-2\sum_{l=1}^Qc_l \sin(lx\delta_x )=0.
\end{equation}
For a certain number $Q$, the above equation is valid if the Taylor expansion of $2\sum_{l=1}^Qc_l \sin(lx\delta_x )$ equals to $x$ until $Q$-$th$ order, and $c_l$ can be obtained consequently.

The quadrupole moment $Q_{xy}$ in the real space\cite{HOD2,HOD3,HOD4} is calculated by:
\begin{equation}
Q_{xy}=\frac{1}{2\pi}Im\{log[det(U^\dagger \widehat{q}U)\sqrt{det(\widehat{q}^\dagger)}]\},
\label{EQ3}
\end{equation}
with $\widehat{q}\equiv e^{i2\pi \widehat{x}\widehat{y}/(N_xN_y)}$. $U$, $\widehat{x}$ and $\widehat{y}$ are the same as those in Eq.~(\ref{EQ2}).

The self-consistent Born approximation \cite{SCBA1} (SCBA) is used to determine the disorder-induced phase transition when disorder strength $W<5t$.
The renormlization self-energy $\Sigma_d$ is obtained by:
\begin{equation}
\Sigma_d=\frac{W^2}{96\pi^3}\iiint_{BZ} dk_xdk_ydk_z[E_F-\mathcal{H}-\Sigma_d]^{-1},
\end{equation}
where $E_F$ the Fermi energy. $W$ represents the disorder strength, and $\mathcal{H}$ is the Hamiltonian in Eq.~(\ref{EQ1}). The integral is on the first Brillouin-zone.

Further, we use the standard transfer matrix method\cite{Scal1,Scal2,Scal3} to determine the metal-insulator transitions. In addition to these numerical approaches, the machine learning method\cite{HOD1,HOD8} is also adopted. More details will be given in section \ref{section4}.

\section{The topological features OF clean HOWSMs}\label{section3}

This section presents key features of the model described by Eq.~(\ref{EQ1}). The difference between the WSMs and HOWSMs is also summarized.

In our calculation, we first adjust the value of the effective mass $m$ and momentum $k_z$ along $z$ direction, and then obtain the Chern number $C$ and quadrupole moment $Q_{xy}$.
In order to determine the phase diagram without disorder, $\mathcal{C}+Q_{xy}\theta(1-\mathcal{C})$ versus $m$ and $k_z$ is plotted in Fig.~\ref{f1}(a), where $\theta(1-\mathcal{C})$ is the Heaviside step function.
 Since the non-zero Chern number will dominant the topological nature of the sample, we use $Q_{xy}\theta(1-\mathcal{C})$ rather than $Q_{xy}$. $Q_{xy}\theta(1-\mathcal{C})$ implies that $Q_{xy}$ is fixed to zero when $\mathcal{C}=1$.
  For $0<|m|<2t$, the Chern number $\mathcal{C}$ and the quadruple moment $Q_{xy}$ both vary with $k_z$.
 Taking $m=0.8t$ as an example, the corresponding $Q_{xy}$ and $\mathcal{C}$ are shown in Fig.~\ref{f1}(b). The quadrupole moment $Q_{xy}$ equals to $0.5$ ($0$) when $\mathcal{C}=0$ ($1$).
 This phase corresponds to the HOWSM\cite{HOWSM2,HOWSM3} because the jump of Chern number ensures the existence of Weyl nodes, and
 $Q_{xy}\theta(1-\mathcal{C})=0.5$ leads to the additional states on hinge.
 For $2t<|m|<4t$, $Q_{xy}\theta(1-\mathcal{C})$ is zero for $k_z\in[-\pi,\pi]$ (see Fig.~\ref{f1}(e) with $m=-2.8t$). Thus, it belongs to the WSM\cite{WSM1} phase instead. As for $|m|>4t$, it is normal insulator (NI) with $\mathcal{C}=0$ and $Q_{xy}=0$. The phase diagram is given in the lower panel of Fig.~\ref{f1}(a).

Next, we present the typical topological features of HOWSMs [see Fig.~\ref{f1}(c)-(d)], and compare them with WSMs [see Fig.~\ref{f1}(f)-(g)]. The first feature of HOWSM is the existence of the hinge states. The $n_{th}$ eigenvalue $E_n$ and the corresponding eigenvector $\psi_n(x,y,z)$ for the above HOWSMs and WSMs are obtained with open boundary along both directions. As illustrated in Fig.~\ref{f1}(c), the HOWSM has obvious zero modes. Moreover, the inverse participation ratio (IPR)\cite{IPR1,IPR2} $I_n=\sum\limits_{x,y,z}|\psi_n(x,y,z)|^4/\sum\limits_{x,y,z}|\psi_n(x,y,z)|^2$ for these modes are higher than others, which suggests that the zero modes are much more localized.
 Further, the corresponding wavefunction $|\psi_n(x,y,z)|^2$ shown in Fig.~\ref{f1}(d) agrees well with the predicted hinge state of HOWSMs (see inset of Fig.~\ref{f1}(d)). In contrast, the IPR for states of the WSM show sharp difference from those in HOWSM, where only the surface states are observable (Fig.~\ref{f1}(f) and (g)).

\begin{figure}[t]
   \centering
    \includegraphics[width=0.48\textwidth]{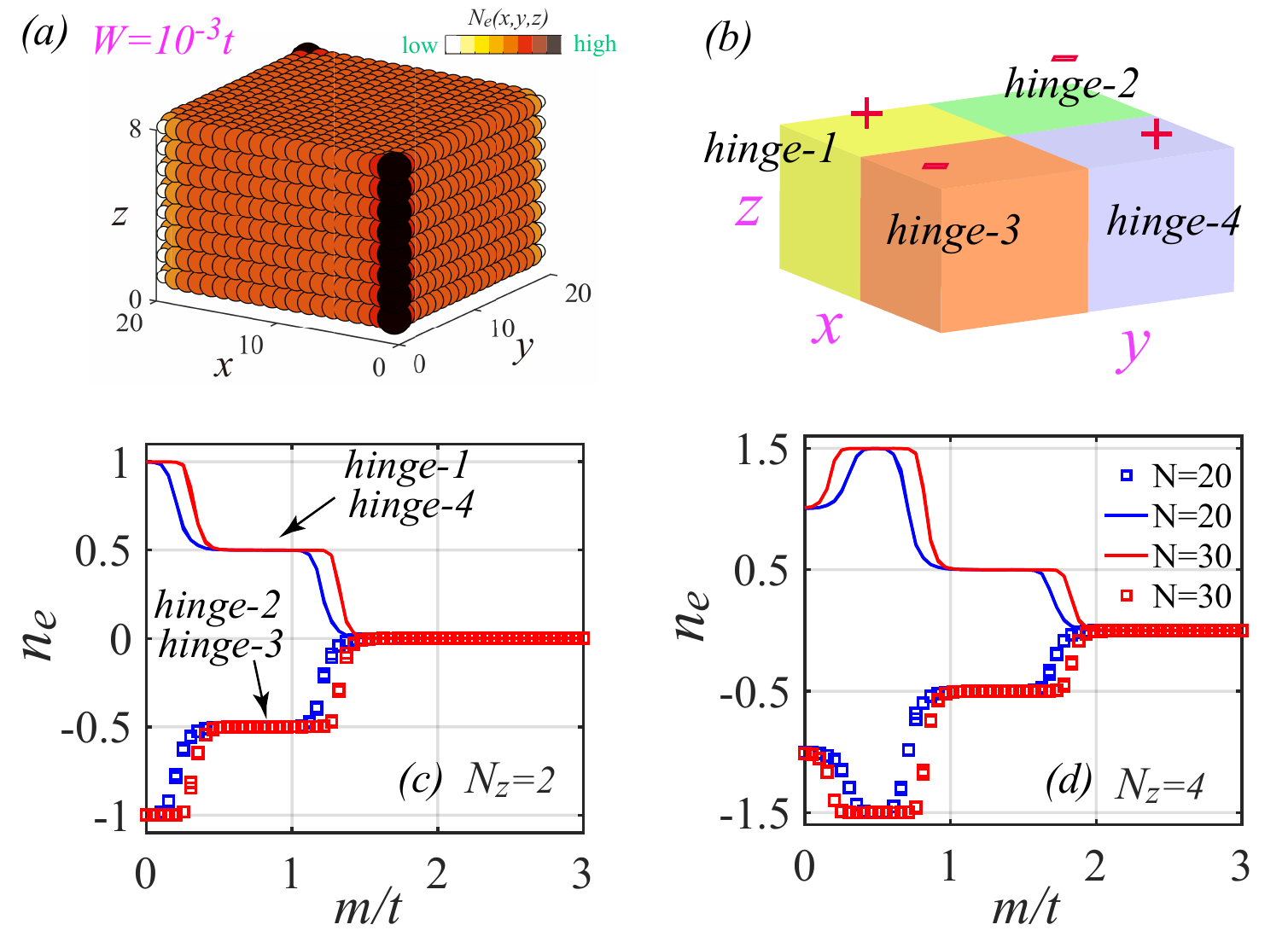}
    \caption{(Color online). (a) The electron density $N_e(x,y,z)$ for HOWSM at half filling with $m=0.8t$ and disorder strength $W=10^{-3}t$. A perturbation with $H_p=\delta(\tau_x\sigma_x+\tau_x\sigma_z)$ is used to lift the degeneracy\cite{HOTI3} between different hinges with $\delta=10^{-3}t$.
    (b) illustrates a distribution for the charge $n_e$ of typical hinges with the sign marked in $\pm$.
    (c) $n_e$ for HOWSM with $N_z=2$ and $N_x=N_y=N$. The hinges with $n_e>0$ and $n_e<0$ correspond to those in (b). (d) The same with (c), except $N_z=4$. }
   \label{f2}
\end{figure}

The second topological feature of HOWSM is the nontrivial quadruple moment $Q_{xy}$. Due to the existence of the Chern number and the $N_z$ dependent of $Q_{xy}$ ($Q_{xy}$ are ambiguous up to {\it mod 1}),
 Eq.~(\ref{EQ3}) is not suitable for HOWSMs with finite size along three directions. Thus, we calculate the electron density distribution for the three-dimensional samples instead. Following the previous works\cite{HOTI3,HOTI4,HOTI5}, we take the half-filling condition. The electron density\cite{HOTI3,HOTI4,HOTI5} is obtained $N_e(x,y,z)=\sum_{n\in occ}{|\psi_n(x,y,z)|^2}$, where {\it occ} stands for the occupied states. To suppress the finite size effect, the periodic boundary condition along $z$ direction is used (except for $N_z=2$). The distribution of $N_e(x,y,z)$ is plotted in Fig.~\ref{f2}(a), in which four hinges construct the quadruple moment distribution.

Further, we have to confirm that each hinge holds the quantized charge, which is also important for HOWSMs.
In the following calculation, the quantized quadruple moment is correlated to the quantized charge instead of $Q_{xy}$ for samples with $N_z>1$.
As shown in Fig.~\ref{f2}(b), we choose the hinge marked in orange as an example.
The charge for such hinge\cite{HOTI3,HOTI4,HOTI5} is $n_e=\sum_{n\in occ}\sum_{z=1}^{N_z}\sum_{x,y=1}^{N/2}[|\psi_{n}(x,y,z)|^2-2]$ with $z$ represents different layer. One needs to delete two positive charges contributed by the atoms under half-filling condition. For a fixed $N_z$, $n_e$ is proportional to the length of the hinge arc in momentum space, which connects two different Weyl-nodes (the length of the solid red line in the inset of Fig.~\ref{f1}(c)). Such a feature is similar to the relationship between the Chern number and the length of the surface arc.

As plotted in Fig.~\ref{f2}(c) with the sample size $N_x\times N_y\times N_z=N\times N\times2$, the fractional quantized charge with $n_e=0.5$ for {\it hinge-1} and {\it hinge-4} is obtained. For {\it hinge-2} and {\it hinge-3}, one has $n_e=-0.5$. Further, with the decrease of $m$ (increase of the length of the hinge arc), $n_e$ gradually moves to $\pm1$.
 Through increasing $N$, the scaling of $n_e$ implies that the nontrivial value is available only when $m<1.5t$ (see Fig.~\ref{f2}(c)), which is different from the phase diagram shown in Fig.~\ref{f1}(a) due to the finite size effect.
To eliminate such a shortage, larger $N_z$ is needed. For $N_z=4$ (Fig.~\ref{f2}(d)), the scaling of $n_e$ suggests the existence of HOWSMs with quantized $n_e$ when $0<m<2t$, in which the finite size effect has been weaken.
In addition, we find $n_e$ increases from $0.5$ to $3\times0.5$ with the decrease of $m$, because the length of the hinge arc increases as expected.
Till now, we have demonstrated the topological features of our model, and the existence of the HOWSMs in the clean limit has been confirmed.

\section{the phase diagram and the fate of the unique topological features of HOWSM}\label{section4}

\begin{figure}[b]
   \centering
    \includegraphics[width=0.38\textwidth]{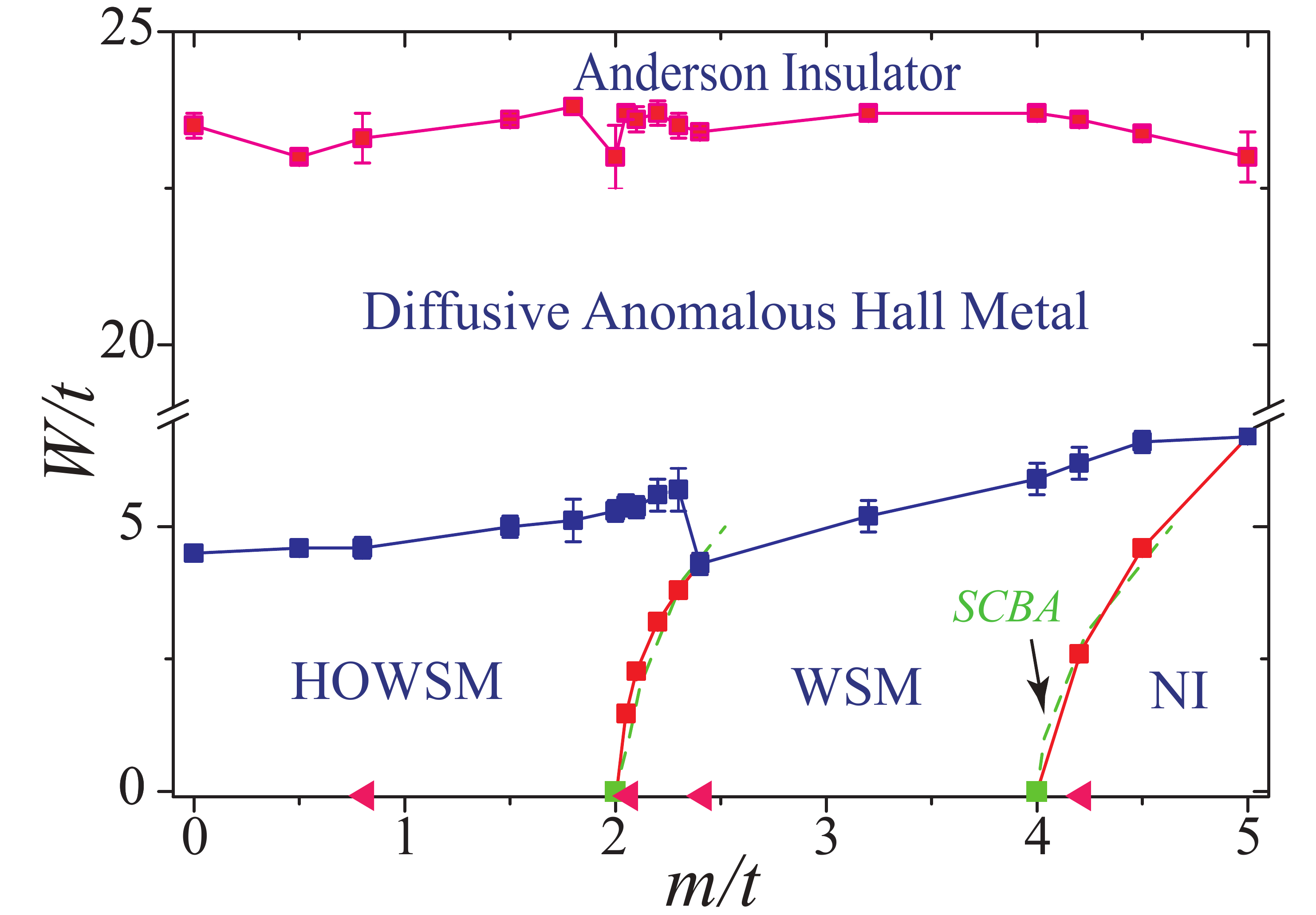}
    \caption{(Color online). The global phase diagram versus ($m$, $W$). When the disorder strength $W=0$, HOWSM, WSM, and NI will appear for different $m$.  The transfer matrix calculation obtains the square points. The green dashed lines are the phase boundaries predicted by SCBA. The red triangles are four typical points. }
   \label{f7}
\end{figure}

 In the following, we focus on the disorder effects of HOWSM.
One of the main results of this paper is given in Fig.~\ref{f7}. It shows the global phase diagram obtained by metal-insulator transition analysis. Our guideline to achieve such a global phase diagram is presented as follows.

\subsection{ Metal-Insulator Transition}

\begin{figure}[t]
   \centering
    \includegraphics[width=0.47\textwidth]{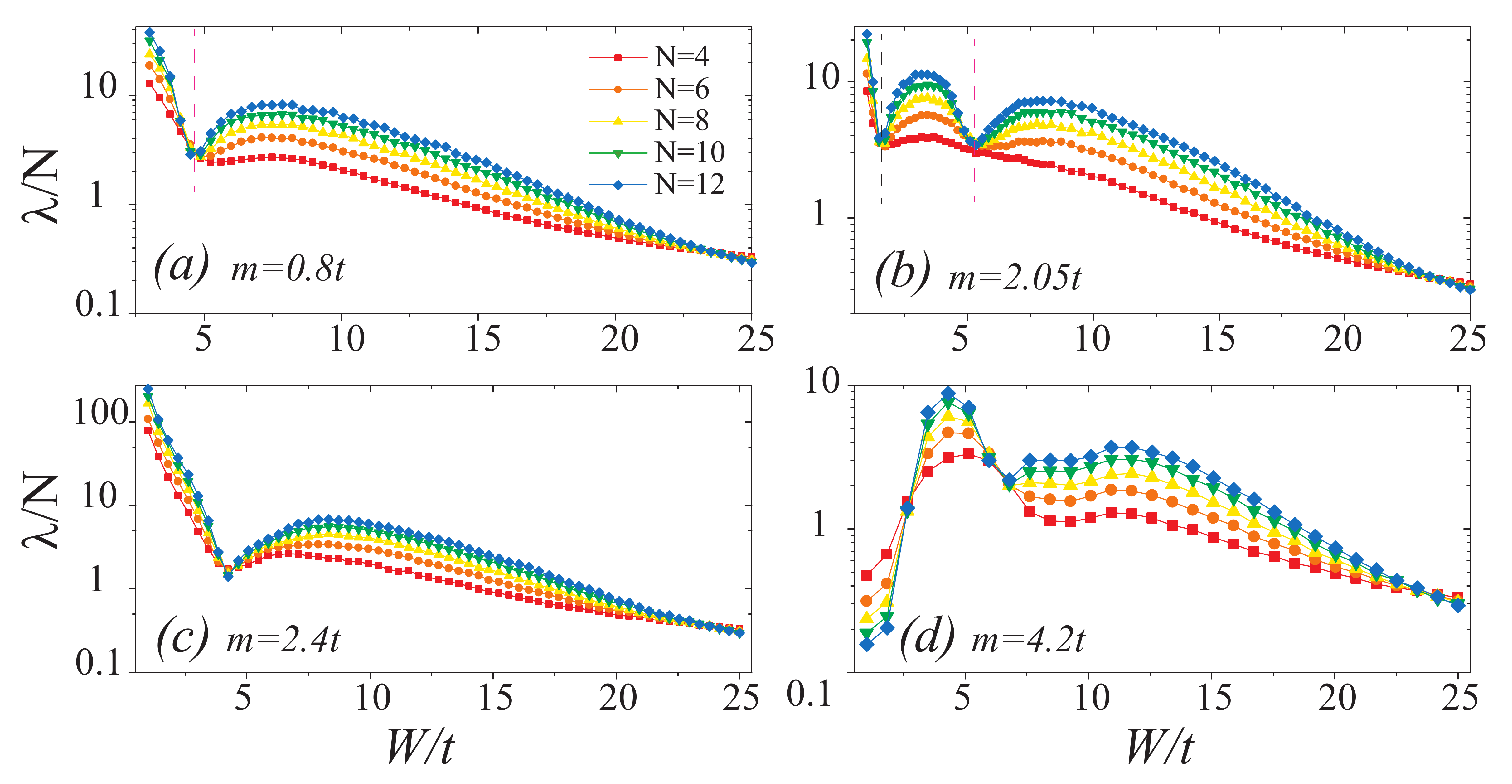}
    \caption{(Color online).  The renormalized localization length $\lambda/N$ versus disorder strength $W$ for different $m$. Four cases are marked in Fig.~\ref{f7} by the red triangles. (a) m=0.8t, (b) $m=2.05t$, (c) $m=2.4t$, (d) $m=4.2t$. The corresponding phases for $W=0$ are (a) HOWSM, (b) WSM, (c) WSM, (d) NI. }
   \label{f3}
\end{figure}

In this subsection, we study the metal-insulator transition of the HOWSMs. When the disorder is absent, our model has the WSM, HOWSM, and NI phases. Notably, the WSM and HOWSM are separated by a single point. We anticipate a disorder-induced phase transition between HOWSM and WSM, and it will be helpful to demonstrate the relationship between WSMs and HOWSMs. The phase transition should be the same for $m>0$ and $m<0$. Thus, we only consider $m>0$ case for simplicity.

Generally, the localization length $\lambda$, which is used to determine the metal-insulator transition, is calculated by the transfer matrix method\cite{WSMD1,WSMD2,Scal1,Scal2,Scal3}. We take the periodic boundary condition along $x$ and $y$ directions and implement the iterative computations along $z$ direction. The sample size along $x$ ($y$) direction is set to $N$. After obtained the localization length $\lambda$, the renormalized localization length $\lambda/N$ is available. There are three cases for the scaling of $\lambda/N$ by increasing sample size $N$. For metallic phases, $\lambda/N$ increases with an increase of $N$.  For insulator phases, $\lambda/N$ decreases with an increase of $N$. For the phase transition point, $\lambda/N$ is unchanged.

Figure.~\ref{f3} plots the typical evolution of $\lambda/N$ versus $W$ for different $m$, which determines the corresponding phases transition of Fig.~\ref{f7}.
 Fig.~\ref{f3}(a) shows a phase transition between two metallic phases when disorder strength $W\approx 4.6t$. A phase transition from metal to insulator occurs when $W\approx 23.3t$. We notice a similar evolution of $\lambda/N$ when $0<m<2t$, where the clean sample belongs to the HOWSM.
On the contrary, for WSMs with $m=2.05t$, we find two-phase transition points among three metallic phases (see Fig.~\ref{f3}(b) the dashed lines). The first transition point between two metallic phases should be different from that shown in Fig.~\ref{f3}(a) because the disorder strength $W\approx1.47t$ is too small compared with $W\approx4.6t$. The rest two transition points are shown in Fig.~\ref{f3}(b), which are corresponding to the two-phase transition points in Fig.~\ref{f3}(a).
 Moreover, the first phase transition point (shown in Fig.~\ref{f3}(b)) shifts to the higher disorder strength with the increase of $m$ and disappears at $m=2.4t$ as shown in Fig.~\ref{f3}(c).
 For $2.4t<m<4t$, the tendency of $\lambda/N$ is similar to those shown in Fig.~\ref{f3}(c). When $m>4t$,
phase transitions insulator$\rightarrow$metal$\rightarrow$metal$\rightarrow$insulator are obtained (see Fig.~\ref{f3}(d)), where the initial clean sample belongs to the NI phase.
We emphasize that two consecutive metal to metal transitions discovered in Fig.~\ref{f3}(b) are distinct, which are rarely reported in Anderson transition before.

\begin{figure}[t]
   \centering
    \includegraphics[width=0.47\textwidth]{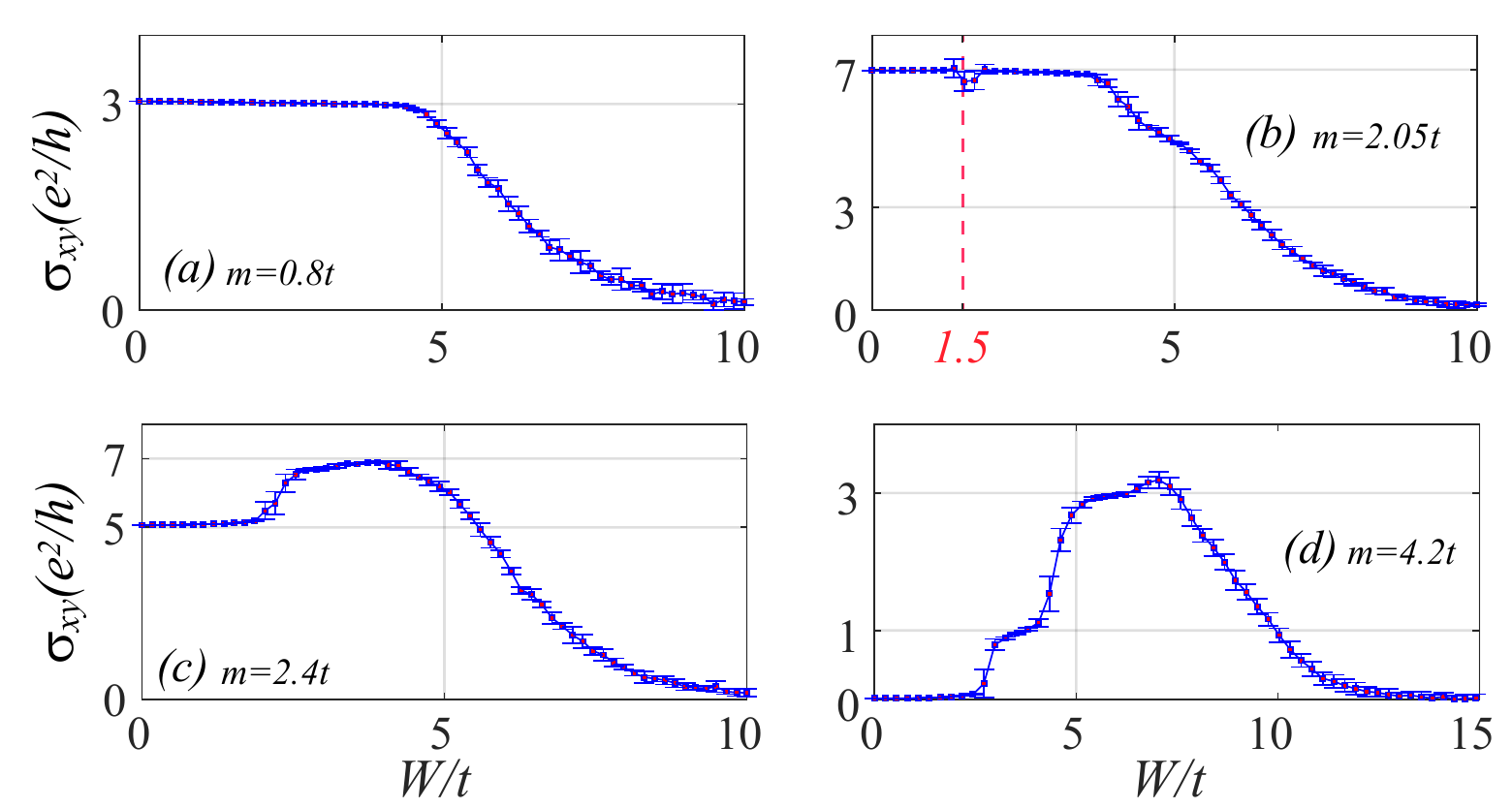}
    \caption{(Color online). The Hall conductance $\sigma_{xy}$ for different $m$. (a) $m=0.8t$, (b) $m=2.05t$, (c) $m=2.4t$, (d) $m=4.2t$. The error bar represents the fluctuation of $\sigma_{xy}$. Sample size is fixed at $20\times20\times8$. The gray line is only for guide. }
   \label{f5}
\end{figure}

 In order to determine the accurate phases of the metallic phases shown in Fig.~\ref{f3}, we now pay attention to the response from the Weyl points.
 As reported before, for a three-dimensional sample with finite-size along both directions, the Chern number is closely related to the distance of two Weyl points\cite{WSMD1}. The Chern number calculation is based on the non-commutative geometry method shown in Eq.~(\ref{EQ2})  with the periodic boundary conditions along three directions adopted. For a sample with size $N_x\times N_y\times N_z=20\times 20\times 8$, we consider all the sites along $z$ direction as the orbit freedom and obtain the Chern number. When the length of the surface arc reaches $2\pi$ ($0$), the Chern number reaches $N_z$ $(0)$.

As shown in Fig.~\ref{f5}(a) ($m=0.8t$), the Chern number is $\mathcal{C}= 3$ with the Hall conductance $\sigma_{xy}=\mathcal{C}e^2/h$. $\mathcal{C}$ is almost unchanged with the increase of disorder strength $W$ and is not equal to $N_z$ (0). Such results imply that the Weyl points are not merged before the disorder destroys it. Further, $\sigma_{xy}$ is neither zero nor quantized for $4.5t<W<10t$.
 Compared with the observations in Fig.~\ref{f3}(a), it is clear that the metal to metal transition here is a phase transition between HOWSM and diffusive anomalous Hall metal (DM).
 As for $W\approx23.3t$, it should be a phase transition between DM and Anderson insulator (AI) with $\sigma_{xy}=0$ (see Fig.~\ref{f7}).
  These results seem like the phase transition results of WSMs\cite{WSMD1,WSMD2,WSMD3,WSMD4,WSMD5,WSMD6}.

 For $m=2.05t$, the quantized Hall conductance is almost unchanged until $W\approx4.5t$ (see Fig.~\ref{f5}(b)). However, there is a dip when $W\approx1.5t$, which corresponds to the first phase transition point shown in Fig.~\ref{f3}(b).
 It may be related to the change of the surface arc's length. Significantly, the first phase transition shown in Fig.~\ref{f3}(b) is a phase transition between the two phases with Weyl-nodes. It should be a phase transition from WSM to HOWSM since the clean sample is a WSM. The second phase transition between two metallic phases in Fig.~\ref{f3}(b) indicates a transition from HOWSM to DM, which is similar to Fig.~\ref{f5}(a).
Therefore, it gives a phase transition between three metals when $m=2.05t$: WSM$\rightarrow$HOWSM$\rightarrow$DM$\rightarrow$AI, which is summarized in Fig.~\ref{f7}. Moreover, the transition from WSM$\rightarrow$ HOWSM is also captured by the SCBA\cite{SCBA1} calculation (the green dashed line shown in Fig.~\ref{f7}). It suggests that such a phase transition originates from the renormalization of the band structure.

 Besides, the phase transition induced by the disorder for $m>2.4t$ can also be confirmed in Fig.~\ref{f5}(c) and (d). One obtains the following phase transitions: WSM$\rightarrow$DM$\rightarrow$AI and NI$\rightarrow$WSM$\rightarrow$DM$\rightarrow$AI. These phase transitions related with WSM are very similar to the previous results reported in WSMs with only two bands\cite{WSMD1,WSMD2,WSMD3,WSMD4,WSMD5,WSMD6}. All the above results are summarized in the global phase diagram shown in Fig.~\ref{f7}.

\subsection{ The stability of the quadrupole moment and the correlated quantized charge}

\begin{figure}[b]
   \centering
    \includegraphics[width=0.47\textwidth]{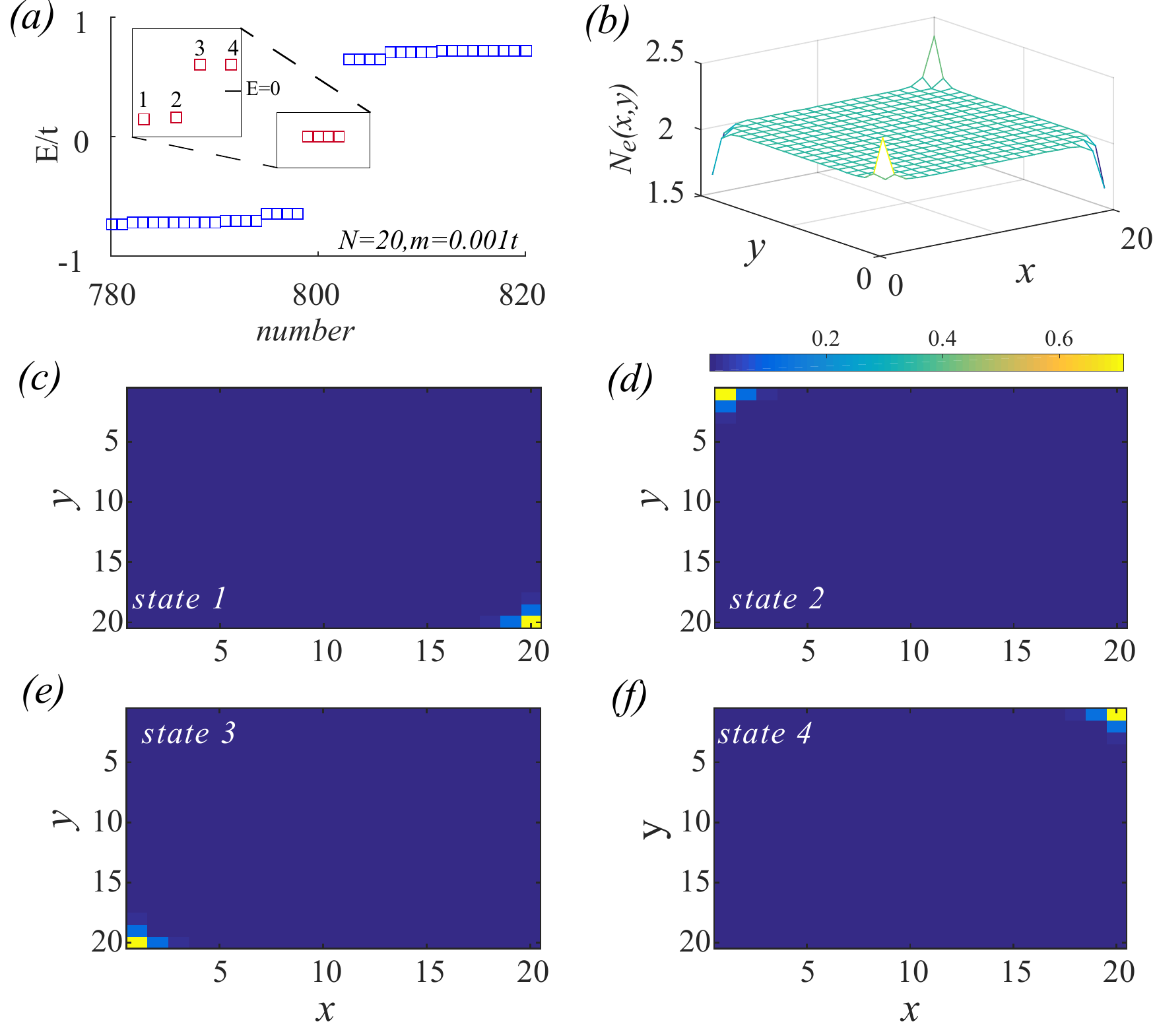}
    \caption{(Color online). (a) The eigenvalues for a HOWSM sample with $N_z=1$ (one layer). (b) The electron density distribution of (a) at half filling. (c)-(f) The eigenstate for the four states marked in red. }
   \label{f4}
\end{figure}

\begin{table}[t]
\centering
\caption{ {\it o/u} implies that the corresponding state is occupied/unoccupied. {\it state-}1 to {\it state-}4 is shown in Fig.~\ref{f4}(c)-(f). The ideal condition for non-trivial $Q_{xy}$ is marked in $\checkmark$. Other cases are marked in $\times$.}\label{TAB1}
\begin{tabular}{l|c|c|c|c|c}
\hline
\hline
 &$state~1$& $state~2$ & $state~3$ &$state~4$   &   $nontrivial~Q_{xy}$\\
 \hline
$case~A$ & {\it o/u} &{\it o/u} & {\it u/o} & {\it u/o}    & $\checkmark$\\
$case~B$ & {\it o/u} &{\it u/o} & {\it u/o} & {\it o/u}     & $\times$   \\
$case~C$ & {\it o/u} &{\it u/o} & {\it o/u} & {\it u/o}      & $\times$   \\
\hline
\hline
\end{tabular}
\end{table}

In the previous subsection, we obtain phase transitions correlated with the HOWSMs. we find that disorder will induce the transition between WSM and HOWSM as well as the transition between HOWSM and DM. Moreover, we prove the stability of Weyl-nodes and the surface arc states for HOWSMs. For instance, as shown in Fig.~\ref{f3}(a) and (b), the Weyl-nodes and the surface arc states can still hold when $W\approx5t$.
Compared to the traditional WSM, the HOWSMs are regarded as possessing the hinge states, the quantized quadrupole moment, and the corresponding quantized charge\cite{HOTI3,HOTI4}.
However, the stability of these additional unique topological features is still unknown. In this section, we study the stability of the quantized quadrupole moment and the corresponding quantized charge of HOWSMs when the disorder effect is considered.

To uncover the stability of the quadrupole moment, we first confine HOWSM samples to one layer. The eigenvalues are plotted in Fig.~\ref{f4}(a), and four states with $E_n\approx0$ are marked in red ($1-4$). These states correspond to the four hinge
(more accurately, they are corners) states ({\it state-}$1$ to {\it state-}$4$) shown in Fig.~\ref{f4}(c)-(f). One needs to consider a small perturbation and lift the fourth-degeneracy of the hinge states\cite{HOTI3,HOTI4}. Then, the electron density distribution shown in Fig.~\ref{f4}(b) at half-filling is available.
As emphasized before\cite{HOTI3,HOTI4}, the specific filling condition and the quantized charge of the hinge states lead to the distribution of Fig.~\ref{f4}(b) and the quantized quadrupole moment.
In concrete, {\it state-}1 and {\it state-}2 are filled (see inset of Fig.~\ref{f4}(a)), which ensures that these two hinges have more electron. On the contrary, the unoccupied {\it state-}3 and {\it state-}4 make these hinges have smaller electron density than other sample areas.

\begin{figure}[t]
   \centering
    \includegraphics[width=0.41\textwidth]{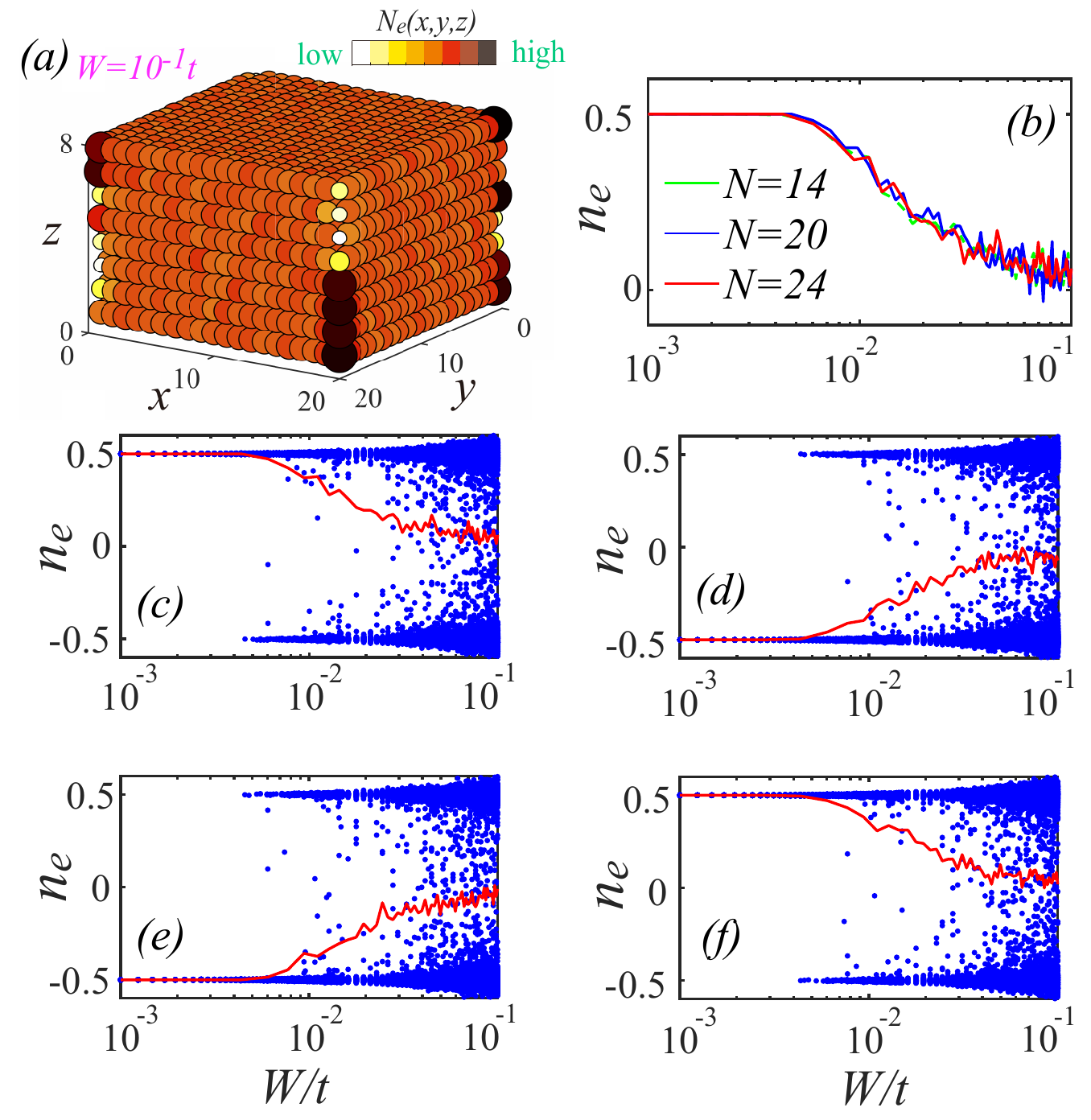}
    \caption{(Color online).
    (a) The spatial distribution of $N_e(x,y,z)$ with disorder strength $W=10^{-1}t$ and $m=0.8t$.
    (b) shows the scaling of the average $n_e$ for different sample sizes.
    We concentrate on {\it hinge-1} since the results for the rest three hinges are similar.
    (c)-(f) The charge $n_e$ versus disorder strength $W$ for the four hinges of the samples, respectively. The red solid line is the average values, and the blue dots represent $n_e$ for each disordered sample. In (b)-(f), we set $m=1.2t$ with sample size $N\times N\times N_z=24\times24\times4$.}
   \label{fcharge}
\end{figure}

When the disorder is considered, however, nothing can guarantee that only {\it state-}1 and {\it state-}2 are filled/unfilled with {\it state-}3 and {\it state-}4 unfilled/filled. The possible filling conditions for the four states, where two of them are filled, are given in TABLE.\ref{TAB1}. Only case A can rebuild an ideal electron density distribution. Thus, the quantized quadrupole moment ensured by four states' special distribution will be destroyed for other cases (even the quantized charge for each hinge still holds).
In $Li's$ paper\cite{HOD4}, the chiral or particle-hole symmetries protect the specific filling condition of the four states.
However, since the the on-site potential induced by the impurity is inevitable in real materials, these symmetries will not available. As shown in Fig.~\ref{fcharge}(a), we present the electron density distribution of a typical HOWSM by considering the Anderson disorder with $W=0.1t$.
Compared with Fig.~\ref{f2}(a), the electron density is no longer regularly distributed.

The second question is the stability of each hinge's quantized charge, which is also important for the quantized quadrupole moment. We still choose samples with four-layer. In section.~\ref{section2}, we have shown that samples with $N_z=4$ capture the quantized charge of HOWSM well. Next, we investigate the evolution of the charge for each hinge $n_e$ with the increase of disorder strength $W$. We take $m=1.2t$ as an example, and over $300$ samples for each disorder strength are counted.
We obtain  $n_e$ for all the samples (the blue dots) and plot how $n_e$ versus with disorder strength $W$ for hinges, shown in Figs.~\ref{fcharge}(c)-(f).
We notice that the blue dots jump between $\pm 0.5$ even when $W\approx 0.007t$. Such observation is consistent with the above discussion, where the required specific filling condition for the quantized quadrupole moment is unstable against disorder.
Further, the solid lines in Fig.~\ref{fcharge}(b)-(f) are the average $n_e$ for different disordered samples.
The quantized average $n_e$ holds only when $W< 0.01t$.
The fluctuation of average $n_e$ increases quickly with an increase of disorder strength $W$.
Moreover, the finite-size scaling of $n_e$ for {\it hinge-1} (see Fig.~\ref{fcharge}(b)) suggests that the quantized charge's stability cannot be improved by increasing the sample size.

Finally, we compare the main findings of this subsection with the previous subsection.
 It is predicted that the HOWSM still holds when $W\approx4.6t$ based on our metal-insulator transition analysis in seciton. \ref{section4} A.
 However, the critical disorder strength, which destroys the quantized charge and the quantized quadrupole moment, is significantly  less than such value.
 Therefore, it is appropriate to conclude that these two topological features of HOWSMs are unstable against disorder.

\subsection{ The fate of the zero-energy hinge states under disorder}

In the previous subsection, we show that the quantized quadrupole moment and quantized charge are unstable and thus challenging to be experimentally observed when the disorder is introduced.
The third feature of HOWSM is the hinge states with wavefunction located at the hinges of the sample (see Fig.~\ref{f1}(d)).
 Next, we study the stability of the hinge modes of HOWSMs and uncover the fate of the disordered HOWSMs.

\begin{figure}[t]
   \centering
    \includegraphics[width=0.48\textwidth]{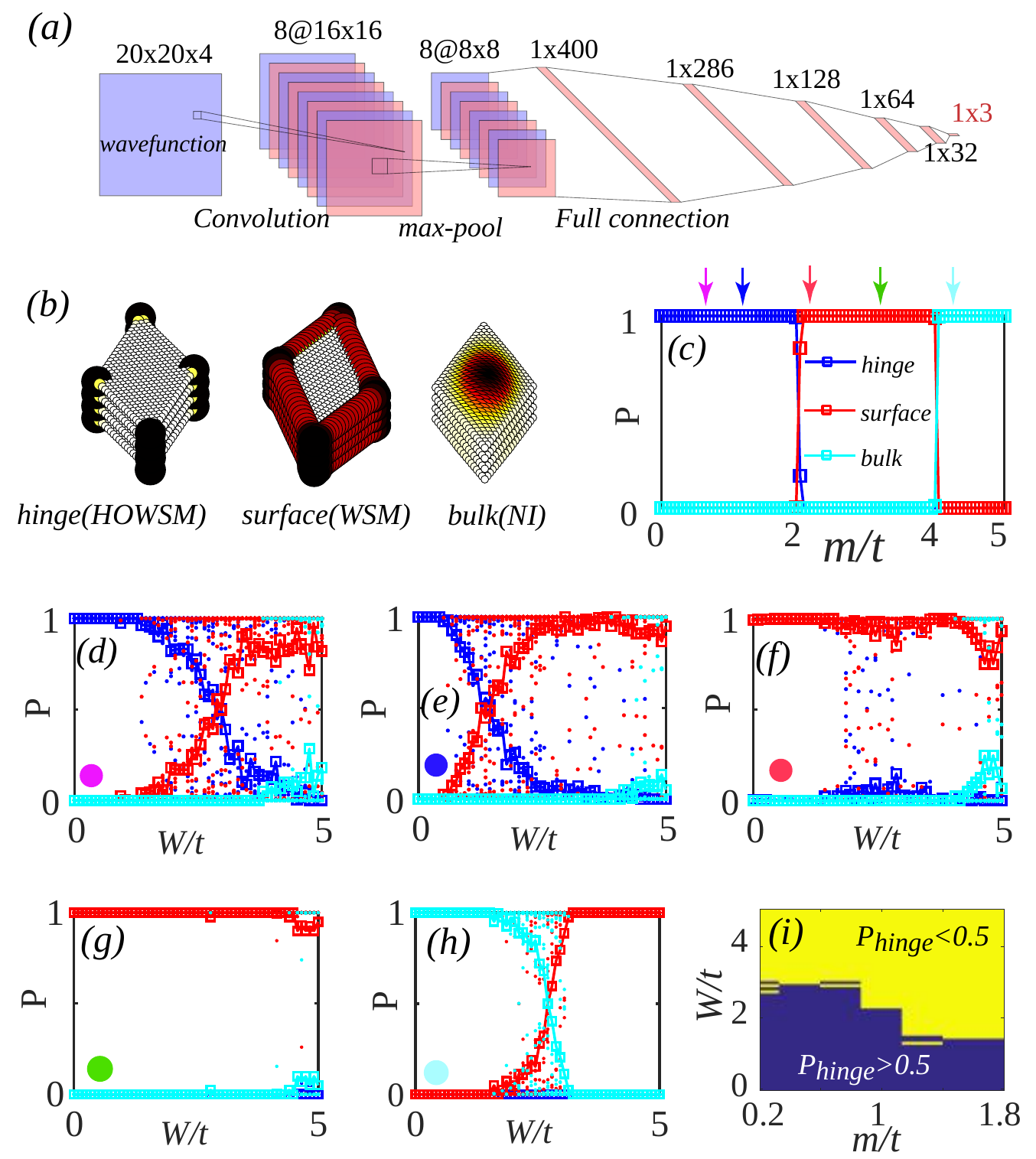}
    \caption{(Color online). (a) The illustration of neural network in calculation. (b) Typical wavefunction distributions for hinge states (HOWSM), surface states (WSM) and bulk states (NI). (c) The probability $\rm{P}$ predicted by the neural network with the evolution of $m$. The arrows are five cases with $m=0.8t$, $m=1.2t$, $m=2.05t$, $m=3.2t$ and $m=4.2t$.  (d)-(h) The evolution of $\rm{P}$ versus disorder strength $W$ with (d) $m=0.8t$, (e) $m=1.2t$, (f) $m=2.05t$, (g) $m=3.2t$, and (h) $m=4.2t$, respectively. The colored dots represent the probability for each sample. The solid lines with square marks are the average probability by counting $40$ disordered samples. (i) represents the probability of existing hinge states ($\rm{P_{hinge}}$) versus $m$ and $W$.  }
   \label{f6}
\end{figure}

Since the traditional methods are inconvenient to determine the stability of hinge modes, the machine learning method is adopted to identify the wavefunction under disorder\cite{HOD1,HOD8}.
 The utilized neural network is shown in Fig.~\ref{f6}(a). It has one convolution layer, one max-pooling layer, and six full-connection layers. TensorFlow 2.0 is used to construct the neural network.
 Since the hinge states are closely related to the zero-energy modes, the wavefunction at half-filling is selected as the input data. We still choose the sample size $20\times20\times4$ and use the periodic boundary condition along $z$ direction (open boundary condition along $x$ and $y$).

There are three kinds of wavefunction distributions for clean samples, shown in Fig.~\ref{f6}(b). The wavefunctions are concentrated at the hinge, surface, and bulk, which correspond to the HOWMS, WSM, and NI, respectively. The training data is obtained based on clean samples with $m\in[0,5t]$. The labels of the training data are determined by distinguishing $m$.
For $m<2t$, it is HOWSM with hinge states. When $m>4t$, it corresponds to the NI with bulk states. For $2t<m<4t$, it is WSM with surface states. As shown in Fig.~\ref{f6}(c), the neural network is trained well since the phase transition points are consistent with the phase diagram presented in Fig.~\ref{f1}(a), and the probability $\rm P$ for the corresponding phases are approximately one.

Because the average wavefunction may bring redundant features that not exist in a single sample, the disordered samples are judged one by one.
 For simplicity, we only show five typical cases, which are marked by the colored arrows in Fig.~\ref{f6}(c) for different $m$.
 The corresponding probability for each sample (marked by the colored dots in Fig.~\ref{f6}(d)-(h)) is investigated, and the average probability for all the disordered samples (solid line with square markers) is also plotted.

In order to check the reliability of the neural network for disordered samples, we give two phase transitions predicted by the neural network.
The first one is shown in Fig.~\ref{f6}(g) with $m=3.2t$, which belongs to the WSM since the surface states with $\rm P\approx1$ hold until $W=5t$. The second one is shown in Fig.~\ref{f6}(h) with $m=4.2t$, and it predicts a transition from bulk states to surface states near $W=3t$, which suggests a phase transition from NI to WSM.
The critical disorder strength and the correlated phase transitions are both consistent with the phase diagram shown in Fig.~\ref{f7} (as well as Fig.~\ref{f3} and \ref{f5}), which preliminarily ensures the reliability of the neural network for disordered samples.

We next study the stability of the hinge states of disordered HOWSMs.
As shown in Fig.~\ref{f6}(d), when $m=0.8t$, the system belongs to the HOWSM under clean limit with $\rm{P_{hinge}}=1$.
$\rm P_{hinge}$ represents the average probability for hinge states and is unchanged when disorder strength satisfies $W<1.2t$. Such critical $W$ means that the hinge states are more robust than the quantized quadrupole moment as well as the quantized charge of hinge states.
We also notice that, if one continues to increase $W$, $\rm P_{hinge}$ decreases sharply and approaches to zero when $W>3.5t$. On the contrary, $\rm P_{surface}$ is zero initially, and it approximately equals to one for $3.5t<W<4.5t$.

The case with $m=0.8t$ is far from the phase boundary ($m=2t$) between WSM and HOWSM in the clean limit.
If $m$ approaches the transition boundary, the hinge states' stability decreases with the main characters unchanged.
It can be verified by the case with $m=1.2t$, shown in Fig.~\ref{f6}(e).
A careful plot of the $\rm{P_{hinge}}$ is given in Fig.~\ref{f6}(i). By increasing $m$ from $0.2t$ to $1.8t$, the critical disorder strength with $\rm{P_{hinge}}\approx 0.5$ decreases from $W_c\approx 2.9t$ to $1.8t$. The decrease of the critical disorder strength is correlated to the decrease of the length of the hinge arc in momentum space. On the other hand, although $W_c$ varies with the variation of $m$, the related disorder strength is still much stronger than the critical strength shown in subsection B. Thus, the hinge states may be detectable in the experiment due to their stability against disorder.

Furthermore, as stated above (see Fig.~\ref{f6}(d)and (e)), the machine learning method predicts that $\rm{P_{surface}}$ can approach to one by increasing $W$ for HOWSMs. This phenomenon seems to signal a disorder induced transition between HOWSM and WSM.
However, such a result is inconsistent with the phase transition determined by transfer matrix, in which there is no phase transition between HOWSMs and WSMs when $0.2t<m<1.8t$ (see Fig.~\ref{f7}, Fig.~\ref{f3} and \ref{f5}).
It should be a fake phase transition and originates from the instability of the hinge states. Nevertheless, the hinge states of HOWSM are more fragile than the Weyl-nodes and the surface arc states.

Due to the hinge states' considerable stability, it is natural to ask whether it can be utilized to characterize the disorder induced phase transition from WSM to HOWSM, shown in Fig.~\ref{f7}.
Taking $m=2.05t$ as an example, the probability for HOWSM and WSM predicted by the neural network are plotted in Fig.~\ref{f6}(f).
 $\rm P_{surface}$ equals to one when $W<1.5t$.
 Then, it slightly decreases with a larger $W$ as $1.5t<W<3.5t$. Meanwhile, $\rm P_{hinge}$ is non-zeros with $\rm{P}\approx[ 0.01,~0.1]$.
 Besides, we notice that some disordered samples have the probability for hinge approximately equals to one (see the blue dots in Fig.~\ref{f6}(f)).
 Such behavior is not observed in Fig.~\ref{f6}(g) and (h), where there is no phase transition correlated with HOWSM.
 It strongly suggests the existence of a phase transition from WSM to HOWSM.
 Nevertheless, $\rm P_{hinge}$ is too small and decreases to zero when $W>3.5t$.
The small value of $\rm P_{hinge}$ is reasonable because the hinge states' length in momentum space is short in this case, and the stability of hinge states is not preserved. In short, the phase transition between WSM and HOWSM may still be difficult to be measured experimentally by using hinge states.

\section{summary  and discussion}\label{section5}

In summary, we studied the disorder effect of HOWSMs.
Firstly, we obtained a global phase diagram, where disorder-induced two metal-metal transitions from WSM to HOWSM as well as HOWSM to DM are observed. These exotic transitions imply that HOWSM is indeed a unique quantum phase, which is different from the WSM and DM.
 Secondly, the fate of the unique topological features of HOWSMs under disorder was checked.
  We found that the extremely weak disorder can destroy the quantized quadrupole moment and the related quantized charge of HOWSMs.
  However, the hinge states are stable under moderate disorder strength.
  For strong disorder, only the Weyl-nodes and the surface arc states still survive.

  Let us discuss the characterization of HOWSM in the experiment. Since the Weyl-nodes are robust enough, it is convenient to check whether the studied sample has the WSM's characters.
  For classical-wave systems, an extremely clean sample can be fabricated. The observation of the quantized quadrupole moment and the related quantized charge will be the ``smoking gun" proof for HOWSMs.
  However, disorder inevitably exists in condensed matter samples, which means the detection of the former two features is impossible. Nevertheless, the hinge states may be observable if the disorder is not too strong, and the HOWSM can be measured.
  Finally, the quantum phase transition usually accompanies with critical behavior of special physical quantities.
   When disorder destroys all the above features of HOWSM, one may still be able to determine the existence of such a phase by detecting the transition between HOWSM and WSM.

\section{acknowledgement}

We are grateful to Yue-Ran Ding, Yan-Zhuo Kang, Qiang Wei, Jie Zhang and Zi-Ang Hu for helpful discussion. This work was supported by National Basic Research Program of China (Grant No. 2019YFA0308403), and NSFC under Grant No. 11822407. C.-Z. C. was funded by the NSFC (under Grant No. 11974256) and the NSF of Jiangsu Province (under Grant No. BK20190813).

\end{document}